\documentstyle[12pt]{article}
\begin{document}
\null\vskip1cm
\centerline{\bf A Lorentz covariant approach to the}
\vskip0.5cm
\centerline{\bf bound state problem}
\vskip1cm
\centerline{L. Micu\footnote{E-mail address:
lmicu@theor1.theory.nipne.ro}}
\vskip0.5cm
\begin{center}
{Department of Theoretical Physics\\
Horia Hulubei Institute for
Physics and Nuclear Engineering\\
 Bucharest POB MG-6, 76900 Romania}
\end{center}
\vskip1cm
\begin{abstract}
The relativistic equivalent of the Schr\"odinger equation for
a two particle bound state having the total angular momentum $S$ is
written in the form of a Lorentz covariant set of equations
$\left(p^\mu_1+p^\mu_2+\Omega^\mu\right)\Psi(p_1,p_2;P)
\chi_S(\vec{p}_1,\vec{p}_2)$=$P^\mu
\Psi(p_1,p_2;P)\chi_S(\vec{p}_1,\vec{p}_2)$  where the operators 
$\Omega^\mu$ are the components of a 4-vector quasipotential.  The
solution of this set is a stationary function representing  the
distribution of spins and internal momenta in a reference frame where
the momentum of the bound system is $P^\mu$. 

The contribution of the operators $\Omega^\mu$ to the bound state
momentum is assumed to be the 4-momentum of a vacuum-like effective
field entering the bound system as an independent component. It is
shown that a state made of free quarks and of the effective field has
definite mass and can be normalized like a single particle state.

The generalization to the case of three or more particles is
immediate.
\end{abstract} 
\newpage

It is a well known fact that the bound state problem has a fully
satisfactory description in classical quantum mechanics, but does not
have a similar one in relativistic field theory. 

In quantum mechanics the existence of a bound state is conditioned by
the presence of a potential well which ensures the localizability
of the wave function in the space of the relative coordinates. In this
context the bound state wave function is stationary and normalizable.

In relativistic field theories the bound state is assumed to be the
result of the continuous exchange of quanta between the constituents
as it is formally expressed by the Bethe-Salpeter method \cite{bs}.
The formalism is fully relativistic, but the number of constituents
in the intermediate states of the iterative solution is indefinite.

It is almost obvious that the connection between the two schemes can
be made only by assuming that the interacting potential in quantum
mechanics is the effective, time averaged result of the continuous
exchange of quanta.  The difference between the two formalisms is
clearly a difference of scale.
One expects then for field theories to be more adequate for the
treatment of high resolution processes occuring at high energy where
the binding effects are negligible, while quantum mechanics to be
suitable for the description of low resolution processes where the
quantum fluctuations cannot be observed as such. It remains still to
find a suitable approach for the bound state to be used in the
treatment of low resolution relativistic processes, where the
binding effects may be significant. 

The approach we propose here is suitable for this last case. It is
based on a covariant generalization of the time independent
Schr\"odinger equation for a bound state. 
This is achieved by replacing the eigenvalue equation
for the  Hamiltonian of the bound system by a set of four eigenvalue
equations for a 4-vector operator representing the sum of the
internal 4-momenta and of a 4-vector interaction potential.
The equations are written in momentum space and their solution 
are the distribution function of the internal spins and momenta.
Lorentz covariance is manifestly satisfied and also is the mass shell
constraint of the bound state momentum.

In the following we refer specifically to QCD and to the meson case
as a quark-antiquark bound state. The generalization to three and
more particles is briefly discussed. The colour indices shall be
omitted for simplicity and the binding potential
will be assumed to be white. 

The equations to be satisfied by the  
distribution function of the internal momenta 
$\Psi_S(p_1,p_2;P)\chi_S(\vec{p}_1,\vec{p}_2$ of a meson with spin $S$
and momentum $P$ with $P^2=M^2$ are:

\begin{equation}\label{set}
\left(~p^\mu_1+p_2^\mu~+~\Omega^\mu \right)
\Psi(p_1,p_2;P)\chi_S(\vec{p}_1,\vec{p}_2)~=~P^\mu~
\Psi(p_1,p_2;P)\chi_S(\vec{p}_1,\vec{p}_2)~~~~~~~\mu=0,1,2,3
\end{equation}
where $p_{1,2}^\mu$ are the on mass shell momenta of the two
quarks. The scalar function $\Psi(p_1,p_2;P)$ represents the
distribution of the quark momenta and
$\chi_S(\vec{p}_1,\vec{p}_2)$ is an expression
involving Dirac spinors, $\gamma$ matrices and quark momenta
having suitable transformation properties. The operators
$\Omega^\mu$ behaving like the components of a 4-vector represent a
relativistic generalization in momentum space of the interaction term
and shall be called a generalized quasipotential.

The generic form of the operators $\Omega^\mu$ can be written with the
aid of the operators  

\begin{equation}
\label{var1} 
i\nabla_{1,2}^\mu=i{\partial\over\partial
p_{1,2\mu}}- i{1\over M^2}\left(P^\alpha{\partial\over\partial
p_{1,2}^\alpha}\right)~P^\mu
\end{equation}
\begin{equation}\label{var2}
p^{T\mu}_{1,2}=p_{1,2}^\mu-{1\over M^2}(P^\alpha
p_{1,2\alpha})~P^\mu  
\end{equation}
\begin{equation}\label{spin1}
\Sigma_{1,2}^\mu~=~~{i\over
M}\epsilon^{\alpha\beta\gamma\mu} P_\alpha \gamma^{(1,2)}_\beta
\gamma^{(1,2)}_\gamma\nonumber
\end{equation}
\begin{equation}\label{spin2}
\Lambda_{1,2}^\mu~=~{i\over M}\epsilon^{\alpha\beta\gamma\mu}
P^\alpha p_{1,2}^\beta \nabla_{1,2}^\gamma
\end{equation} 
representing the relativistic covariant generalizations of the
position, momentum, spin and angular momentum in the rest frame of
the meson. We have then: 

\begin{equation}\label{omega} 
\Omega^\mu = {P^\mu\over
M}~\tilde{\mathcal U}_0~+~[p^{T\mu}_1,~\tilde{\mathcal
U}_{1}]_+~+~[p^{T\mu}_2, ~\tilde{\mathcal U}_{2}]_+ + i[\nabla_1^\mu,
 ~\tilde{\mathcal V}_{1}]_+ + i[\nabla_2^\mu, ~\tilde{\mathcal
V}_{2}]_+ 
\end{equation}
where $\tilde{\mathcal V},~\tilde{\mathcal U}$ are scalar
hermitian operators. We notice that ${\mathcal U}_{0,1,2}$ are even at
time reversal, while ${\mathcal V}_{1,2}$ are odd. 

In the rest frame of the meson the equations (\ref{set})
take the form: 

\begin{equation}\label{restset1}
\left(\sqrt{\vec{p}_1^2+m_1^2}+\sqrt{\vec{p}_2^2+m_2^2}
+\tilde{\mathcal U}_0\right)
\Psi(p_1,p_2;M)\chi_S(\vec{p}_1,\vec{p}_2)
=M~\Psi(p_1,p_2;M)\chi_S(\vec{p}_1,\vec{p}_2)
\end{equation} 
\begin{equation}\label{restset2}
\left(\vec{p}_1+\vec{p}_2+[\vec{p}_1,\tilde{\mathcal
U}_1]_++[\vec{p}_2,\tilde{\mathcal U}_2]_++
i[\nabla_1,\tilde{\mathcal V}_1]_++i[\nabla_2~\tilde{\mathcal
V}_2]_+\right) \Psi(\vec{p}_1,\vec{p}_2)\chi_S(\vec{p}_1,\vec{p}_2)
=0.  
\end{equation}

If $\vec{\Omega}=0$ in this frame the solution of (\ref{restset2})
reads 

\begin{equation}\label{sol}
\Psi(p_1,p_2;M)~
=~\delta^{(3)}(\vec{p}_1+\vec{p}_2)~ \psi(\vec{p})\chi_S(\vec{p}).
\end{equation}
In addition, if $\tilde{\mathcal U}_0$ depends on
$\vec{p}_1-\vec{p}_2$ only, in the nonrelativistic limit 
eq.(\ref{restset1}) reduces to the Fourier transform of the usual
Schr\"odinger equation for a bound state 

\begin{equation}\label{sol1}
({\vec{p}^2\over 2\mu}~+~
\tilde{\mathcal U}_0) \psi(\vec{p})\chi_S(\vec{p})~=~
(M-m_1-m_2)~\psi(\vec{p})\chi_S(\vec{p}), 
\end{equation}
where $\mu={m_1~m_2\over
m_1+m_2}$ is the reduced mass
and $\vec{p}={1\over2}(\vec{p}_1-\vec{p}_2)$ is half of the relative
quark momentum in the rest frame of the bound state. 

We notice also that
the operators $\Lambda_1$ and $\Lambda_2$ 
are now replaced by the orbital angular momentum $\vec{L}$ of the
relative motion 
\begin{equation}
L_k~=~-i\epsilon_{ijk}p^i~\nabla_p^j\nonumber 
\end{equation} 
and the spin-spin and the spin-orbit couplings are 
$\vec{\sigma}_{1,2}\cdot\vec{\sigma}_{1,2}$ and
$\vec{\sigma}_{1,2}\cdot\vec{L}$ where $\sigma_i$ are Pauli matrices. 

Then, just like in quantum mechanics one can
write the orthogonality relation for the solutions of the eigenvalue
equation (\ref{sol1}):

\begin{equation}\label{norm1}
\int d^3p~
\psi_M(\vec{p})~\psi_{M'}^*(\vec{p})~=\delta_{MM'}. 
\end{equation} 

Equations (\ref{sol1}) and (\ref{norm1}) have been written for
completeness and also for clarifying the relation between
the set (\ref{set}) and the Schr\"odinger equation for a bound state.

Taking advantage of the relativistic covariance of (\ref{set}) and
(\ref{restset1},\ref{restset2}) one may solve them in the rest frame
and write their solutions in any other frame simply by
replacing any scalar product $\vec{p}_i \cdot \vec{p}_j$ by its
Lorentz covariant expression  $-p_i^{T\mu}\cdot p_{j\mu}^T={1\over
M^2} (P\cdot p_i)~(P\cdot p_j)- (p_i\cdot p_j)$. 

We recall that equation (\ref{restset1}) with the solution
(\ref{sol}) and ${\mathcal U}_0$ a relativistic scattering kernel
\cite{qp}, \cite{njl} and/or a suitable confining potential
\cite{fm}, \cite{gi}, \cite{gm}  represent the starting point for
most relativistic quark models. In all these cases dynamics is
restricted to the relative coordinate. 
Our approach is more general. The relative motion does not have a
special significance and the quarks can be really treated as
independent particles. This
makes a generalization of the set (\ref{set}) to the baryon  case
very easy by including the contribution of a third quark momentum and
writing the suitable relativistic coupling of spins and angular
momenta.

The relation between the Bethe Salpeter
formalism and the present approach also becomes clear by looking
at their solutions written as perturbative series with respect to the
interaction term. In the BS formalism the solution writes
with the aid of relativistic Green functions having poles at positive
and negative energies. This makes the number of particles in the
intermediate states to be indefinite and causes a lot of trouble.
In our case the perturbative series must be written for the
relativistic invariant equation 

\begin{equation}
(P_\mu p_1^\mu+P_\mu p_2^\mu+P_\mu
\Omega^\mu)\Psi(p_1,p_2;P) \chi_S(\vec{p}_1,\vec{p}_2)=M^2
\Psi(p_1,p_2;P) \chi_S(\vec{p}_1,\vec{p}_2)
\end{equation}
which has been derived from the set (\ref{set}) and coincides with
(\ref{restset1}) in the meson rest frame. Its Green function is
Lorentz invariant and has only positive energy poles. This makes
the perturbative series to resemble the series in 
nonrelativistic quantum mechanics and the number of particles to be
the same in the external and intermediate states.  

We analyse now the way one can use the
solutions of the set (\ref{set}) to obtain a suitable representation
of a meson in terms of independent quark operators. It is obvious
that a meson is not an ensemble of free quarks with the momentum
distribution $\Psi_S(p_1,p_2;P)$  because the sum of the
free quark 4-momenta does not coincide with the bound state
4-momentum. In order to have a real dynamical representation of the
bound state the contribution of the operators $\Omega^\mu$ to the
meson momentum must also be included. 

The solution we found to this problem is to write the
contribution of the operators $\Omega^\mu$ as
the 4-momentum of an independent component of the meson existing
besides the valence quarks \cite{micu}. To this end we introduce the
notation   
\begin{eqnarray}\label{Q}
&&\Omega^\mu
\Psi(p_1,p_2;P)\chi_S(\vec{p}_1,\vec{p}_2)
=(P^\mu-p_1^\mu-p_2^\mu)~  \int d^4Q~\varphi(p_1,p_2;Q)
~\sum_{s_1,s_2}\bar u_{s_1}(\vec{p}_1)\Gamma_S
v_{s_2}(\vec{p}_2) \nonumber\\ 
&&= \int d^4Q~Q^\mu~\varphi(p_1,p_2;Q)~
\sum_{s_1,s_2}\bar
u_{s_1}(\vec{p}_1)\Gamma_S v_{s_2}(\vec{p}_2). 
\end{eqnarray}  
where $\Gamma_S$ is a suitable tensor product of $\gamma$ matrices and
quark momenta. The single meson state is written then as follows:

\begin{eqnarray}\label{meson}
&&\left.\vert M_i(P)\right\rangle=
\int d^3p_1~{m_1\over e_1}
d^3p_2{m_2\over e_2}d^4Q 
~\delta^{(4)}(p_1+p_2+Q-P)\nonumber\\
&&\times\varphi(p_1,p_2;Q)
\sum_{s_1,s_2}\bar u_{s_1}(\vec{p}_1)\Gamma_S
v_{s_2}(\vec{p}_2)~\Phi^\dagger(Q)~
 a^\dagger_{s_1} (\vec{p}_1)b^\dagger_{s_2}
(\vec{p}_2)\vert 0\rangle 
\end{eqnarray}
where  $a^\dagger$ and $b^\dagger$ are quark and
antiquark creation operators and $u,v$ are free Dirac
spinors. $\Phi^\dagger$ is a vacuum-like effective field carrying the
momentum  $Q^\mu=P^\mu-p_1^\mu-p_2^\mu$ defined in the relation
(\ref{Q}). According to its definition the field $\Phi$ represents
the collective, time averaged effect of all the virtual
excitations responsible for the binding and as such it
must be compared with the "bag" in bag models \cite{bm}. We notice
however that the bag has positive potential energy characterized
by a constant volume density $B$, while the 
effective field $\Phi$ carries the 4-momentum $Q^\mu$ whose values
are {\it a priori} limited only by the quark and meson mass shell
constraints. (see eq. (\ref{meson}).) The existence of a momentum
$\vec{Q}$ besides the energy component $Q_0$ may be considered the
consequence of the imperfect cancellation of the vector momenta in the
virtual processes generating the binding and is in agreement with
the opinion that the potential is the average effect of a continuous
series of quantum fluctuations. 

The relativistic equivalent of eq.(\ref{norm1}) is now the
orthonormality relation for the single meson state (\ref{meson}). 
We calculate it in the general case by using the
commutation relations for the free quark operators and
assuming the following expression for the vacuum expectation value of
the effective field:

\begin{equation}\label{vev}
\left\langle 0\right\vert~\Phi(Q_1)~\Phi^+(Q_2)~\left\vert 0 
\right\rangle~={1\over L^3_0T_0}~
\int d^4~X~{\rm e}^{i~(Q_2-Q_1)_\mu~X^\mu}=
{(2\pi)^4\over L^3_0T_0}~\delta^{(4)}(Q_1-Q_2)
\end{equation}
where $L_0$ is the range of the binding forces and $T$ is a time
sensibly larger than the time basis involved in the definition of the
effective field.

We notice that 
$\delta(Q_0-Q'_0)$ in (\ref{vev}) induces a cumbersome $\delta(E-E')$
in the expression of the norm. In order to avoid it by preserving the
manifest Lorentz covariance of eq. (\ref{vev}) we write

\begin{equation}\label{delta} 
{2\pi\over T_0}~\delta(Q_0-Q'_0)
={1\over T_0}\int_T~dX_0~{\rm e}^{i(E(P)-E(P'))X_0}
\approx {E\over MT_0}\int_{T_0}
dY_0 {\rm e}^{i(M-M')Y_0}\approx{E\over M}~\delta_{MM'}
\end{equation}
and get immediately:

 \begin{equation}\label{norm}
\left\langle~{\cal M'}(P')~\vert {\cal
M}(P)~\right\rangle~=
2E~(2\pi)^3~\delta^{(3)}(P-P')~\delta_{MM'}~{\cal J}
\end{equation}
where

\begin{eqnarray}\label{J2} 
&&{\cal J}=~{1\over 2M~L_0^3}~\int d^3p_1~{m_1\over
e_1}~d^3p_2~{~m_2\over e_2}~d^4Q~
\delta^{(4)}(p_1+p_2+Q-P)\vert \varphi(p_1,p_2;Q)\vert^2\nonumber\\
&&\times Tr\left({\hat p_1+m_1\over 2m_1}~\Gamma_S~
{-\hat p_2+m_2\over2m_2}~\Gamma_{S'}\right)=1.
\end{eqnarray}

In the above relations we have implicitly assumed that $M$ and
$M'$ are discrete eigenvalues of the  equation (\ref{restset1}) with
$\vert M-M'\vert~T_0>>1$ and hence the integral in (\ref{delta})
vanishes if $M\ne M'$. Notice the remarkable fact that ${\mathcal
J}$ does not depend on the rather arbitrary time $T_0$. 

In the same way as above it can be shown that the expression of the 
norm will contain the highly singular factor
$\delta(\alpha-\alpha')\delta^3(0)$ if $\vec{\Omega}$
vanishes in (\ref{restset1}) i.e. if $Q^\mu=\alpha P^\mu$ where
$\alpha$ is a scalar. 

Relation (\ref{norm}) shows that the bound state
function of the many particle  system representing the meson can be
normalized like that of a single particle if the integral $\cal J$
converges. 

We expect for the internal function $\varphi(p_1,p_2;P)~\bar u_{s_1}
(\vec{p}_1)\Gamma_S v_{s_2}(\vec{p}_2)$ which represents an
equilibrium distribution  of spins and momenta in the bound system to
have a stationary counterpart in the coordinate space which must
coincide in the nonrelativistic limit with the internal wave function
of the meson. Proceeding like in the BS formalism we define the meson
wave function in coordinate space

\begin{equation}\label{wf}
\tilde\varphi(\vec{x},\vec{y},\vec{X},t;P)= \langle 0\vert 
~\bar\psi(\vec{x},t)\Gamma_{S'}
\psi(\vec{y},t)\Phi(\vec{X},t)\vert{\cal
M}(\vec{P},E)\rangle 
\end{equation}
where the single meson state is given by (\ref{meson}), $\psi$ is
the quark field

\begin{equation}\label{psi}
\psi(\vec{x},t)={1\over (2\pi)^3}\sum_{s}~\int~d^3p~{m\over
e_p}\left(b^\dagger_{\vec{p},s}~v(\vec{p},s)~{\mathrm
e}^{ie_pt-\vec{p}\vec{x}}+ a_{\vec{p},s}~u(\vec{p},s) ~{\mathrm
e}^{-ie_pt+i\vec{p}\vec{x}}\right)
\end{equation}
and 
\begin{equation}
\Phi(\vec{X},t;Q_0)={\sqrt{L^3_0T_0}\over(2\pi)^2}\int~d^4Q {\mathrm
e}^{-iQ_0t+\vec{Q}\vec{X}} \Phi(\vec{Q},Q_0).
\end{equation}

After a straightforward calculation we get 

\begin{equation}\label{wf1}
\tilde\varphi(\vec{x},\vec{y},\vec{X},t;P)=
{\mathrm
e}^{-iEt+i\vec{P}\cdot\vec{X}}~\tilde\phi(\vec{x},\vec{y},\vec{X};P)
\end{equation}
where
\begin{eqnarray}\label{wf2}
&&\tilde\phi(\vec{x},\vec{y},\vec{X};P)=
{(2\pi)^2\over\sqrt{L^3_0T_0}} 
\int~{d^3p_1\over 2e_1}{d^3p_2\over 2e_2}\nonumber\\
&&\times Tr\left(\Gamma_S(\hat
p_1+m_1)\Gamma_{S'}(-\hat
p_2+m_2)\Gamma_S)\right)\varphi(p_1,p_2;P-p_1-p_2){\mathrm
e}^{i\vec{p}_1\cdot (\vec{x}-\vec{X})}{\mathrm
e}^{i\vec{p_2}\cdot(\vec{y}-\vec{X})}
\end{eqnarray}
describes the internal meson structure in coordinate space.  

At the first sight the expression (\ref{wf2}) does not
coincide with the usual internal wave function of a bound state
because of its dependence on the coordinate $\vec{X}$ associated to
the effective field which has no classical equivalent. However, 
in the nonrelativistic limit $\tilde\phi$ writes as:

\begin{equation}
\tilde\phi(\vec{x},\vec{y},\vec{X};P)=
\delta^{(3)}(\vec{X}-{\vec{x}+\vec{y}\over2})\tilde\phi_0
(\vec{x},\vec{y};P)
\end{equation}
if $\varphi$ depends on the relative momentum $\vec{p}_1-\vec{p}_2$
only. Here $\phi_0$ is really the internal wave function in
nonrelativistic quantum mechanics because it represents the Fourier
transform of the nonrelativistic momentum distribution function.

Concluding this paper we remark that the present approach
provides a relativistic nonperturbative solution of the bound
state problem. The main differences from the
older relativistic approaches are the covariant
generalization of the time independent Schr\"odinger equation
accompanied by the introduction of a 4-vector quasipotential and the
the definition of an effective field as an additional, independent
constituent of the bound system besides the valence quarks. The
4-momentum of the effective field represents the
contribution of the interaction term to the hadron momentum. The
solutions of the dynamical equations (\ref{set}) define equilibrium
distributions of quark spins and momenta inside hadrons and hence may
be directly used to calculate electroweak formfactors in
the soft limit where binding effects are important.  
\vskip0.5cm    
{\bf Acknowledgements}
This work was initiated during author's visit in 1998 at the
Institute of Theoretical Physics of the University of Bern.  The
author thanks Prof. Heiri Leutwyler for hospitality and stimulating
discussions.

\end{document}